# Reviving the past: the growth of citations to old documents

**Alberto Martín-Martín[1], Enrique Orduña-Malea[2], Juan Manuel Ayllón[1], Emilio Delgado López-Cózar[1]**

[1] *EC3: Evaluación de la Ciencia y de la Comunicación Científica, Universidad de Granada (Spain)*
[2] *EC3: Evaluación de la Ciencia y de la Comunicación Científica, Universidad Politécnica de Valencia (Spain)*

**A DIGEST OF**

Verstak A, Acharya A, Suzuki H, Henderson S, Lakhiaev M, Chiung Yu Lin C, Shetty N. On the Shoulders of Giants: The Growing Impact of Older Articles.
**Full article available from:** http://arxiv.org/abs/1411.0275

## SUMMARY

In this *Digest* we review a recent study released by the Google Scholar team on the apparently increasing fraction of citations to old articles from studies published in the last 24 years (1990-2013). First, we describe the main findings of their article. Secondly, we conduct an analogue study, using a different data source as well as different measures which throw very similar results, thus confirming the phenomenon. Lastly, we discuss the possible causes of this phenomenon.

**1. DIGEST**

| |
|---|
| **RESEARCH QUESTIONS** |
| § How often are older articles cited in scholarly papers and how has this changed over time? <br> § How does the impact of older articles vary across different fields of scholarship? <br> § Is the change in the impact of older articles accelerating or slowing down? <br> § Are these trends different for much older articles? |
| **METHODOLOGY** |
| **Unit analysis** |
| Citations from English articles published in scientific journals and conferences (1990-2013), indexed in the 2014 release of Google Scholar Metrics |
| **Sample** |
| Citations from English articles published in scientific journals and conferences (1990-2013), indexed in the 2014 release of Google Scholar Metrics |
| **Design** |
| § This study covers English scientific journals and conferences assigned to one or more subject categories (261) from the 2014 release of Google Scholar Metrics. <br> § The 261 subject categories are grouped into 9 broad research areas. <br> § For each journal and conference, all articles with a publication date within 1990-2013 are considered. <br> § For each category-year/area-year group, the total number of citations as well as the number of citations to articles published in each preceding year is computed. <br> § Three different thresholds for older articles were used: ≥ 10 years old; ≥ 15 years old; and ≥ 20 years old. <br> § To see if the rate of change in the fraction of older citations is speeding up or slowing down, the aggregate change for 1990-2001 (first half) and 2002-2013 (second half) for every category are computed. |
| **Measures** |
| § Percentage of citations to older articles (articles that were published at least 10, 15 and 20 years before the citing article) from articles published in English scientific journals, indexed in the 2014 release of Google Scholar Metrics. <br> § Percentage of citations to older articles for 9 broad areas of research. <br> § Rate of change in the fraction of older citations: the aggregate change for 1990-2001 (first half) and 2002-2013 (second half) for every category are computed. |
| **Period analyzed:** 1990-2013 |
| **Data collection date:** Unknown, but data from the 2014 edition of Google Scholar Metrics (which was released in June 2014) is used. |

**RESULTS**

1. In 1990, 28% of citations were to articles that were at least 10 years old. In 2013, this percentage was 36%, a growth of 28% since 1990 (Table 1).
2. The fraction of older citations increased over 1990-2013 for 7 out of 9 broad areas of research and 231 out of 261 subject categories (Table 1).

**Table 1: Change in the fraction of older citations over 1990-2013**

| Broad area | Old citations 1990 % | Old citations 2013 % | Change 1990-2013 % |
|---|---|---|---|
| Humanities, Literature & Arts | 44 | 51 | 18 |
| Business, Economics & Management | 30 | 46 | 56 |
| Social Sciences | 33 | 43 | 31 |
| Physics & Mathematics | 31 | 40 | 29 |
| Life Sciences & Earth Sciences | 29 | 39 | 36 |
| Engineering | 33 | 34 | 3 |
| Chemical & Material Sciences | 32 | 33 | 2 |
| Health & Medical Sciences | 25 | 33 | 30 |
| Computer Science | 20 | 28 | 39 |
| All articles | 28 | 36 | 28 |

3. This growth occurs in almost all scientific disciplines: 102 out of 261 subject categories saw a growth in the fraction of older citations that was over 30%, 44 of them with an increase over 50%. For Business, Economics & Management and Computer Science, almost two-thirds of the subject categories saw a growth over 50%. On the other hand, in Chemical & Material Sciences and Engineering, most of the subject categories show a drop in the fraction of citations to old articles.
4. The change over the second half (2002-2013) was significantly larger than over the first half (1990-2001) — the increase in the second half was double the increase in the first half.

**Table 2. Change in the fraction of citations to older articles over 1990-2001 & 2002-2013**

| Broad area | Change over 1990-2001 | Change over 2002-2013 |
|---|---|---|
| Humanities, Literature & Arts | 4% | 14% |
| Business, Economics & Management | 19% | 37% |
| Social Sciences | 5% | 26% |
| Physics & Mathematics | 11% | 18% |
| Life Sciences & Earth Sciences | 11% | 25% |
| Engineering | 2% | 1% |
| Chemical & Material Sciences | -1% | 3% |
| Health & Medical Sciences | 4% | 26% |
| Computer Science | 8% | 31% |
| All articles | 9% | 19% |

Baseline for all growth percentages is the fraction of older citations in 1990.

5. The trend of a growing impact of older articles also holds for articles that are at least 15 years old and those that are at least 20 years old. If in 1990, 16% of citations were to articles ≥ 15 years old and 10% to articles ≥ 20 years old, in 2013 these figures rose to 21% and 13% respectively, with growth rates higher than 30%.

## 2. DISCUSSION

With an attractive and suggestive title, this work brings up interesting and relevant questions. After defining a set of clear and precise goals, the authors describe a simple and straightforward methodological design - very adequate to answer these questions -, and lastly they present clear and convincing results.

However, the methods section should have indicated the exact size of the object of study: the number of journals, articles and citations that have been processed.

Thus, there are some questions that remain unanswered:

- How many journals do they refer to when they say "all the categorized journals and conferences, not only the top 20 per category"?
- How many articles do they refer to when they say "articles with a publication date within 1990-2013"?
- How many citations do they refer to when they say "total citations as well as the number of citations for each preceding year, included all the citations from these articles"?
- How many citations have been processed in total?
- Where are the results for each of the 261 subject categories? Why aren't they included as a table or Appendix?
- Why don't they offer the raw data so it can be analyzed by other researchers?

On another note, it is important to stress that these results refer to journals written in English. Would the results be different if journals written in other languages had been analyzed instead? Ruiz & Jiménez (1996) discovered, for a sample of Library and Information Science journals, two different paces of aging: one for English written journals, and one for the rest of journals.

In order to check if the results shown in this work can be confirmed using other data sources that cover journals written in languages other than English, and at the same time using alternative procedures to calculate the pace of aging of citations, we decided to replicate this study. To do this, we have used data from Thomson Reuters' Journal Citation Reports.

Since Gross & Gross (1927) introduced the concept of obsolescence, that is, the phenomenon by which scientific publications are decreasingly used over time, various methods have been proposed to measure the aging process of scientific literature.

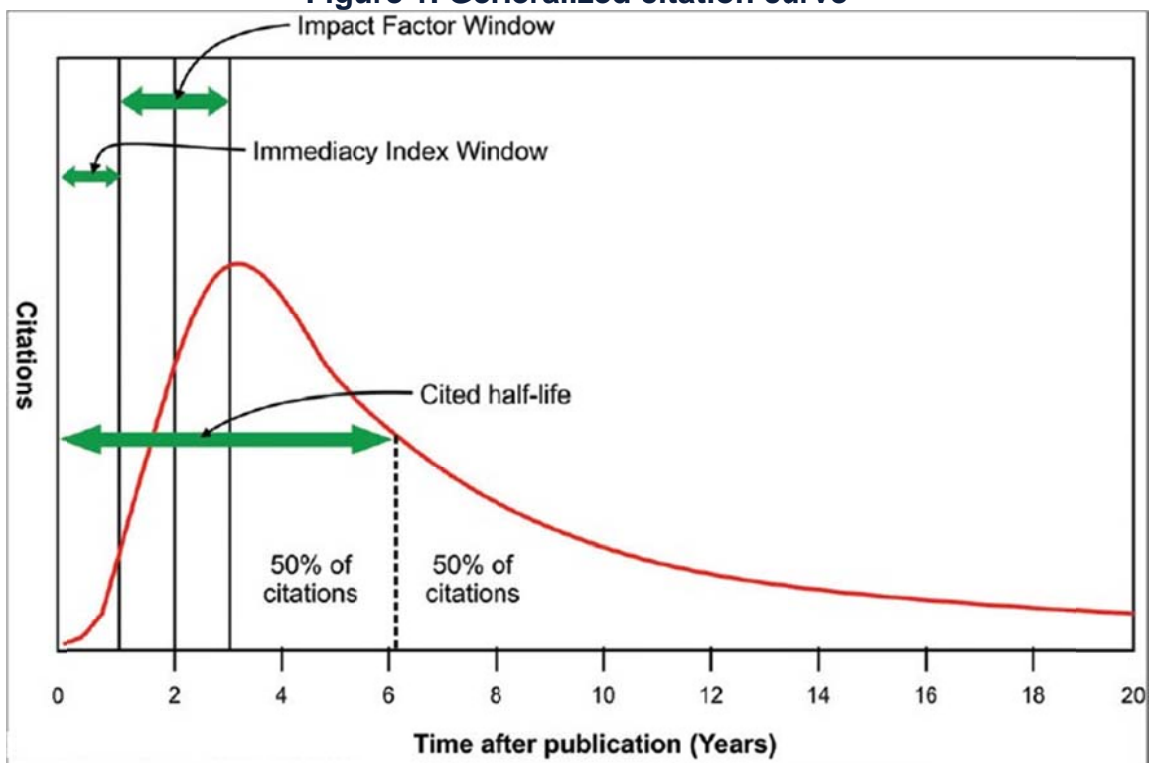
**Figure 1. Generalized citation curve**

Source: Elsevier Science

Ruiz & Bailón (1998) empirically analyzed five methods, comparing and assessing the quality of the results and the statistical errors each of these methods presented. Among these measures, the most popular one is known by the name of "Half-Life", and was first proposed by Burton & Kebler in 1960.

Our study uses the "Cited Half-Life" indicator, used by Thomson Reuters, defined as "the median age of the articles that were cited in Journal Citation Reports each year". Every year, Thomson Reuters calculates an "Aggregate Cited Half-Life" for the 53 subject categories present in the Social Sciences edition of the JCR, and also for the 167 Science categories.

In order to explain this indicator, we offer the following example: in JCR 2003, the subject category Energy & Fuels has a cited half-life of 7.0. This means that articles published in Energy & Fuels journals between 1997-2003 (inclusive) account for 50% of all citations to articles from those journals in 2003. Since the first time the JCR included the Aggregate Cited Half-Life was in 2003, we have used this year and compared it to the data shown in JCR 2013.

Therefore, we can't accurately replicate the study by Verstak et al., which analyzes data from 1990 onwards. However, we'll be able to observe the evolution of this last decade (2003-2013). Moreover, another limitation in our study is that we have not been able to analyze journals from the Arts & Humanities. There is no JCR for A&H, and therefore their Cited Half-Life indicators have not been calculated.

Journal, article and citation data used in the calculation of the cited half-life for the 220 subject categories present in the JCR 2003 and 2013 are presented in Table 3, whereas the aggregate cited Half-Life for these subject categories, and the change in that decade is shown in Table 4.

Since there are many journals indexed in more than one subject category, the sum of these elements is not equal to the number of unique journals, articles and citations (Table 3).

**Table 3. Absolute number of Journals, Articles and Citations covered in the Journal Citation Reports (2003 and 2013)**

| JCR | Citations | Journals | Articles |
|---|---|---|---|
| 2013 | 74,835,917 | 17,638 | 2,292,350 |
| 2003 | 29,973,394 | 11,710 | 1,240,712 |

**Table 4. Aggregate Cited Half-Life subject categories Journal Citations Reports 2003 and 2013**

| Subject Categories | 2003 | 2013 | Change | Trend | Subject Categories | 2003 | 2013 | Change | Trend |
|---|---|---|---|---|---|---|---|---|---|
| History | >10,0 | >10,0 | | ? | Planning & Development | 7,0 | 8,7 | 1,7 | ▲ |
| History & Philosophy of Science | >10,0 | >10,0 | | ? | Ergonomics | 8,4 | 8,6 | 0,2 | ▲ |
| History of Social Sciences | >10,0 | >10,0 | | ? | Criminology & Penology | 7,9 | 8,5 | 0,6 | ▲ |
| Social Sciences, Mathematical Methods | >10,0 | >10,0 | | ? | Communication | 8,3 | 8,4 | 0,1 | ▲ |
| Sociology | >10,0 | >10,0 | | ? | Social Work | 8,0 | 8,4 | 0,4 | ▲ |
| Psychology, Educational | >10,0 | >10,0 | | ? | Women's Studies | 7,2 | 8,4 | 1,2 | ▲ |
| Psychology, Mathematical | >10,0 | >10,0 | | ? | Psychiatry | 7,7 | 8,3 | 0,6 | ▲ |
| Psychology, Social | >10,0 | >10,0 | | ▲ | Urban Studies | 6,5 | 8,3 | 1,8 | ▲ |
| Economics | >10,0 | >10,0 | | ? | Education & Educational Research | 8,2 | 8,2 | 0,0 | = |
| Business | 9,9 | >10,0 | | ▲ | Social Sciences, Interdisciplinary | 8,5 | 8,1 | -0,4 | ▼ |
| Psychology, Multidisciplinary | 9,7 | >10,0 | | ▲ | Gerontology | 6,5 | 8,1 | 1,6 | ▲ |
| Psychology, Applied | 9,5 | >10,0 | | ▲ | International Relations | 6,5 | 7,9 | 1,4 | ▲ |
| Business, Finance | 9,4 | >10,0 | | ▲ | Public Administration | 7,2 | 7,8 | 0,6 | ▲ |
| Psychology, Psychoanalysis | 9,3 | >10,0 | | ▲ | Social Issues | 7,1 | 7,7 | 0,6 | ▲ |
| Management | 9,1 | >10,0 | | ▲ | Transportation | 7,6 | 7,4 | -0,2 | ▼ |
| Law | 8,5 | >10,0 | | ▲ | Geography | 6,7 | 7,4 | 0,7 | ▲ |
| Industrial Relations & Labor | 8,4 | >10,0 | | ▲ | Information Science & Library Science | 6,1 | 7,4 | 1,3 | ▲ |
| Psychology, Biological | 9,3 | 9,7 | 0,4 | ▲ | Ethics | 7,4 | 7,3 | -0,1 | ▼ |
| Anthropology | 9,6 | 9,6 | 0,0 | = | Rehabilitation | 7,1 | 7,2 | 0,1 | ▲ |
| Area Studies | 7,6 | 9,3 | 1,7 | ▲ | Public, Environmental & Occupational Health | 7,0 | 7,2 | 0,2 | ▲ |
| Demography | 8,4 | 9,2 | 0,8 | ▲ | Social Sciences, Biomedical | 6,9 | 7,2 | 0,3 | ▲ |
| Political Science | 8,3 | 9,2 | 0,9 | ▲ | Substance Abuse | 6,3 | 7,0 | 0,7 | ▲ |
| Psychology, Developmental | 8,8 | 9,0 | 0,2 | ▲ | Health Policy & Services | 6,0 | 7,0 | 1,0 | ▲ |
| Psychology, Experimental | 9,0 | 8,8 | -0,2 | ▼ | Education, Special | 8,8 | 6,9 | -1,9 | ▼ |
| Psychology, Clinical | 8,2 | 8,8 | 0,6 | ▲ | Nursing | 6,9 | 6,9 | 0,0 | ▲ |
| Family Studies | 8,2 | 8,7 | 0,5 | ▲ | Ethnic Studies | 6,1 | 6,8 | 0,7 | ▲ |
| | | | | | Environmental Studies | 7,2 | 6,5 | -0,7 | ▼ |

| Subject Categories | 2003 | 2013 | Change | Trend | Subject Categories | 2003 | 2013 | Change | Trend |
|---|---|---|---|---|---|---|---|---|---|
| Mathematics | >10,0 | >10,0 | | ? | Physiology | 7,2 | 8,7 | 1,5 | ▲ |
| Statistics & Probability | >10,0 | >10,0 | | ? | Mining & Mineral Processing | 7,1 | 8,7 | 1,6 | ▲ |
| Mineralogy | >10,0 | >10,0 | | ? | Mathematics, Applied | 8,7 | 8,6 | -0,1 | ▼ |
| Ornithology | >10,0 | >10,0 | | ? | Ecology | 8,4 | 8,6 | 0,2 | ▲ |
| Zoology | >10,0 | >10,0 | | ? | Agricultural Economics & Policy | 8,2 | 8,6 | 0,4 | ▲ |
| Limnology | 9,7 | >10,0 | | ▲ | Physics, Mathematical | 6,7 | 8,6 | 1,9 | ▲ |
| Agriculture, Soil Science | 9,6 | >10,0 | | ▲ | Operations Research & Management Science | 9 | 8,5 | -0,5 | ▼ |
| Geochemistry & Geophysics | 9,5 | >10,0 | | ▲ | Agriculture, Dairy & Animal Science | 8,8 | 8,5 | -0,3 | ▼ |
| Paleontology | 8,9 | >10,0 | | ▲ | Otorhinolaryngology | 8,6 | 8,5 | -0,1 | ▼ |
| Engineering, Aerospace | 8,6 | >10,0 | | ▲ | Biology | 7,7 | 8,5 | 0,8 | ▲ |
| Geology | 8,3 | >10,0 | | ▲ | Engineering, Petroleum | >10,0 | 8,4 | | ▼ |
| Mathematics, Interdisciplinary Applications | >10,0 | 9,9 | | ▼ | Dentistry, Oral Surgery & Medicine | 8,9 | 8,4 | -0,5 | ▼ |
| Entomology | 9,4 | 9,7 | 0,3 | ▲ | Forestry | 7,7 | 8,4 | 0,7 | ▲ |
| Engineering, Geological | 9,3 | 9,7 | 0,4 | ▲ | Geosciences, Multidisciplinary | 7,7 | 8,4 | 0,7 | ▲ |
| Oceanography | 8,5 | 9,7 | 1,2 | ▲ | Behavioral Sciences | 8,1 | 8,3 | 0,2 | ▲ |
| Fisheries | 8,4 | 9,6 | 1,2 | ▲ | Evolutionary Biology | 7,3 | 8,3 | 1,0 | ▲ |
| Marine & Freshwater Biology | 8,7 | 9,5 | 0,8 | ▲ | Physics, Fluids & Plasmas | 6,3 | 8,3 | 2,0 | ▲ |
| Computer Science, Theory & Methods | 8,2 | 9,3 | 1,1 | ▲ | Sport Sciences | 8,2 | 8,2 | 0,0 | = |
| Horticulture | 7,7 | 9,3 | 1,6 | ▲ | Imaging Science & Photographic Technology | 6,7 | 8,2 | 1,5 | ▲ |
| Acoustics | 8,3 | 9,2 | 0,9 | ▲ | Nuclear Science & Technology | 6,5 | 8,2 | 1,7 | ▲ |
| Physics, Atomic, Molecular & Chemical | 8,9 | 9,1 | 0,2 | ▲ | Biochemistry & Molecular Biology | 6,0 | 8,2 | 2,2 | ▲ |
| Mechanics | 9,8 | 9,0 | -0,8 | ▼ | Microscopy | 5,6 | 8,2 | 2,6 | ▲ |
| Anatomy & Morphology | 7,6 | 9,0 | 1,4 | ▲ | Developmental Biology | 5,4 | 8,2 | 2,8 | ▲ |
| Engineering, Ocean | 6,6 | 8,9 | 2,3 | ▲ | Engineering, Industrial | 7,5 | 8,1 | 0,6 | ▲ |
| Orthopedics | 9,3 | 8,7 | -0,6 | ▼ | Medicine, General & Internal | 7,0 | 8,1 | 1,1 | ▲ |
| Materials Science, Ceramics | 8,2 | 8,7 | 0,5 | ▲ | Veterinary Sciences | 7,8 | 8,0 | 0,2 | ▲ |
| Agronomy | 8,1 | 8,7 | 0,6 | ▲ | Agriculture, Multidisciplinary | 7,5 | 8,0 | 0,5 | ▲ |
| Computer Science, Hardware & Architectur | 8,1 | 8,7 | 0,6 | ▲ | Engineering, Mechanical | 7,4 | 8,0 | 0,6 | ▲ |
| Plant Sciences | 7,5 | 8,7 | 1,2 | ▲ | Physics, Multidisciplinary | 7,1 | 8,0 | 0,9 | ▲ |
| Physiology | 7,2 | 8,7 | 1,5 | ▲ | Anesthesiology | 6,9 | 8,0 | 1,1 | ▲ |

| Subject Categories | 2003 | 2013 | Change | Trend | Subject Categories | 2003 | 2013 | Change | Trend |
|---|---|---|---|---|---|---|---|---|---|
| Materials Science, Paper & Wood | 9,5 | 7,9 | -1,6 | ▼ | Rehabilitation | 7,2 | 7,3 | 0,1 | ▲ |
| Computer Science, Software Engineering | 7,9 | 7,9 | 0,0 | ▲ | Clinical Neurology | 6,8 | 7,3 | 0,5 | ▲ |
| Ophthalmology | 7,4 | 7,9 | 0,5 | ▲ | Food Science & Technology | 7,4 | 7,2 | -0,2 | ▼ |
| Remote Sensing | 7,1 | 7,9 | 0,8 | ▲ | Emergency Medicine | 6,6 | 7,2 | 0,6 | ▲ |
| Multidisciplinary Sciences | 6,9 | 7,9 | 1,0 | ▲ | Mycology | 6,3 | 7,2 | 0,9 | ▲ |
| Computer Science, Cybernetics | 6,7 | 7,9 | 1,2 | ▲ | Substance Abuse | 6,0 | 7,2 | 1,2 | ▲ |
| Surgery | 7,6 | 7,8 | 0,2 | ▲ | Respiratory System | 5,6 | 7,2 | 1,6 | ▲ |
| Psychiatry | 6,9 | 7,8 | 0,9 | ▲ | Cell Biology | 5,4 | 7,2 | 1,8 | ▲ |
| Materials Science, Coatings & Films | 6,8 | 7,8 | 1,0 | ▲ | Metallurgy & Metallurgical Engineering | 6,9 | 7,1 | 0,2 | ▲ |
| Peripheral Vascular Disease | 5,6 | 7,8 | 2,2 | ▲ | Thermodynamics | 8,2 | 7,0 | -1,2 | ▼ |
| Medical Laboratory Technology | 7,7 | 7,7 | 0,0 | = | Transportation Science & Technology | 7,6 | 7,0 | -0,6 | ▼ |
| Meteorology & Atmospheric Sciences | 7,3 | 7,7 | 0,4 | ▲ | Robotics | 7,3 | 7,0 | -0,3 | = |
| Pathology | 6,8 | 7,7 | 0,9 | ▲ | Obstetrics & Gynecology | 6,8 | 7,0 | 0,2 | ▲ |
| Physics, Nuclear | 6,8 | 7,7 | 0,9 | ▲ | Polymer Science | 6,8 | 7,0 | 0,2 | ▲ |
| Neurosciences | 6,2 | 7,7 | 1,5 | ▲ | Chemistry, Organic | 6,7 | 7,0 | 0,3 | ▲ |
| Biophysics | 5,9 | 7,7 | 1,8 | ▲ | Engineering, Electrical & Electronic | 6,7 | 7,0 | 0,3 | ▲ |
| Water Resources | 7,6 | 7,6 | 0,0 | = | Computer Science, Information Systems | 6,6 | 7,0 | 0,4 | ▲ |
| Dermatology | 7,2 | 7,6 | 0,4 | ▲ | Medical Informatics | 6,4 | 7,0 | 0,6 | ▲ |
| Spectroscopy | 6,2 | 7,6 | 1,4 | ▲ | Radiology Nuclear Medicine & Medical Imaging | 6,3 | 7,0 | 0,7 | ▲ |
| Geography, Physical | 7,3 | 7,5 | 0,2 | ▲ | Astronomy & Astrophysics | 6,2 | 7,0 | 0,8 | ▲ |
| Chemistry, Inorganic & Nuclear | 7,0 | 7,5 | 0,5 | ▲ | Pharmacology & Pharmacy | 6,2 | 7,0 | 0,8 | ▲ |
| Computer Science, Artificial Intelligence | 6,5 | 7,5 | 1,0 | ▲ | Endocrinology & Metabolism | 6,0 | 7,0 | 1,0 | ▲ |
| Reproductive Biology | 5,8 | 7,5 | 1,7 | ▲ | Toxicology | 6,0 | 7,0 | 1,0 | ▲ |
| Biodiversity Conservation | 9,2 | 7,4 | -1,8 | ▼ | Engineering, Manufacturing | 5,9 | 7,0 | 1,1 | ▲ |
| Construction & Building Technology | 7,5 | 7,4 | -0,1 | ▼ | Genetics & Heredity | 5,4 | 7,0 | 1,6 | ▲ |
| Pediatrics | 7,0 | 7,4 | 0,4 | ▲ | Immunology | 5,4 | 7,0 | 1,6 | ▲ |
| Microbiology | 5,8 | 7,4 | 1,6 | ▲ | Engineering, Marine | >10,0 | 6,9 | | ▼ |
| Critical Care Medicine | 5,2 | 7,4 | 2,2 | ▲ | Nursing | 7,0 | 6,9 | -0,1 | ▼ |
| Crystallography | 8,3 | 7,3 | -1,0 | ▼ | Materials Science, Characterization & Testing | 6,8 | 6,9 | 0,1 | ▲ |
| Engineering, Multidisciplinary | 7,6 | 7,3 | -0,3 | ▼ | Andrology | 6,7 | 6,9 | 0,2 | ▲ |

| Subject Categories | 2003 | 2013 | Change | Trend | Subject Categories | 2003 | 2013 | Change | Trend |
|---|---|---|---|---|---|---|---|---|---|
| Chemistry, Analytical | 6,7 | 6,9 | 0,2 | ▲ | Neuroimaging | 4,9 | 6,4 | 1,5 | ▲ |
| Tropical Medicine | 7,8 | 6,8 | -1,0 | ▼ | Biochemical Research Methods | 6,8 | 6,3 | -0,5 | ▼ |
| Nutrition & Dietetics | 6,4 | 6,8 | 0,4 | ▲ | Engineering, Biomedical | 6,2 | 6,3 | 0,1 | ▲ |
| Chemistry, Applied | 6,2 | 6,8 | 0,6 | ▲ | Allergy | 5,7 | 6,3 | 0,6 | ▲ |
| Materials Science, Composites | 6,2 | 6,8 | 0,6 | ▲ | Medicine, Legal | 5,3 | 6,3 | 1,0 | ▲ |
| Computer Science, Interdisciplinary Applica | 6,0 | 6,8 | 0,8 | ▲ | Optics | 6,6 | 6,2 | -0,4 | ▼ |
| Health Care Sciences & Services | 6,0 | 6,8 | 0,8 | ▲ | Oncology | 5,4 | 6,2 | 0,8 | ▲ |
| Medicine, Research & Experimental | 6,0 | 6,8 | 0,8 | ▲ | Transplantation | 5,3 | 6,2 | 0,9 | ▲ |
| Cardiac & Cardiovascular Systems | 5,8 | 6,8 | 1,0 | ▲ | Infectious Diseases | 5,2 | 6,2 | 1,0 | ▲ |
| Automation & Control Systems | 7,8 | 6,7 | -1,1 | ▼ | Telecommunications | 6,5 | 6,1 | -0,4 | ▼ |
| Physics, Condensed Matter | 6,7 | 6,7 | 0,0 | ▲ | Physics, Particles & Fields | 4,8 | 6,1 | 1,3 | ▲ |
| Environmental Sciences | 6,5 | 6,7 | 0,2 | ▲ | Electrochemistry | 6,7 | 6,0 | -0,7 | ▼ |
| Urology & Nephrology | 5,6 | 6,7 | 1,1 | ▲ | Physics, Applied | 6,0 | 6,0 | 0,0 | ▲ |
| Engineering, Civil | 8,0 | 6,6 | -1,4 | ▼ | Engineering, Environmental | 6,2 | 5,9 | -0,3 | ▼ |
| Geriatrics & Gerontology | 5,8 | 6,6 | 0,8 | ▲ | Integrative & Complementary Medicine | 5,5 | 5,9 | 0,4 | ▲ |
| Engineering, Chemical | 7,4 | 6,5 | -0,9 | ▼ | Chemistry, Physical | 5,4 | 5,7 | 0,3 | ▲ |
| Materials Science, Textiles | 7,4 | 6,5 | -0,9 | ▼ | Chemistry, Multidisciplinary | 7,2 | 5,6 | -1,6 | ▼ |
| Virology | 5,7 | 6,5 | 0,8 | ▲ | Medical Ethics | 5,7 | 5,5 | -0,2 | ▼ |
| Hematology | 5,4 | 6,5 | 1,1 | ▲ | Materials Science, Multidisciplinary | 5,6 | 5,4 | -0,2 | ▼ |
| Biotechnology & Applied Microbiology | 5,3 | 6,5 | 1,2 | ▲ | Parasitology | 6,9 | 5,2 | -1,7 | ▼ |
| Instruments & Instrumentation | 6,1 | 6,4 | 0,3 | ▲ | Materials Science, Biomaterials | 5,6 | 5,1 | -0,5 | ▼ |
| Chemistry, Medicinal | 6,0 | 6,4 | 0,4 | ▲ | Agricultural Engineering | 7,7 | 5,0 | -2,7 | ▼ |
| Rheumatology | 5,8 | 6,4 | 0,6 | ▲ | Energy & Fuels | 7,0 | 4,7 | -2,3 | ▼ |
| Gastroenterology & Hepatology | 5,7 | 6,4 | 0,7 | ▲ | | | | | |

The results shown in Table 4 are clear:

- On average, the half-life of citations in journals increases from 7.2 years to 7.5. This is true both for Science & Technology journals (from 7 to 7.4) and Social Sciences journals (from 7.9 to 8.1).
- This growth is present across the majority of subject categories: there is an increase in 159 categories, a decrease in 42 categories, and in 6 categories the indicator stays the same. There are also 13 categories for which we ignore this information.
- The disciplines with a higher cited half-life are those linked to the Humanities (History, Philosophy) and the Social Sciences (Sociology, Economics & Business, Psychology). Within Science & Technology, Mathematics, the biological sciences (Zoology, Ornitology), and the Geosciences have also a high cited half-life.
- The number of disciplines with a cited half-life higher than 10 years in 2013 is twice what it was back in 2003.
- The subject categories with a higher increase in their cited half-life are quite diverse (Developmental Biology, Microscopy, Engineering Ocean, Peripheral Vascular Disease, Critical Care Medicine, Biochemistry & Molecular Biology, Physics: Fluids & Plasmas, Physics Mathematical).

The subject categories where there has been a higher decrease in their cited half-life belong mostly to engineering and chemistry, specially Material Science: Energy & Fuels, Agricultural Engineering, Special Education, Biodiversity Conservation, Parasitology, Chemistry Multidisciplinary, Materials Science: Paper & Wood.

As can be seen (Table 4), the results offered in the Journal Citation Reports / Web of Science match very closely the results described by Verstak et al., which have been reached using Google Scholar data.

That said, the reasons that explain this extension in the life cycle of scientific documents are still open to discussion.

The first factor that should be considered has already been studied extensively, and it is the relation between the exponential growth of scientific production and the pace of obsolescence. In 1963, Price suggested this bond, although it was Line, in 1970, who described with more detail the relationship between these two phenomena, determining that if the number of published articles grows rapidly, an equally rapid growth in the number of citations to recently published articles can be expected. He confirmed that the faster the pace at which scientific studies are published, the faster these publications become obsolete and stop being cited.

However, in 1993, Egghe mathematically proved how the growth of scientific production modifies the pace of obsolescence, pointing out the technical differences between diachronous and synchronous studies. He confirmed how obsolescence increases in synchronous studies (like the one carried out by Verstak et al.) and decreases in diachronous studies. In various brainy as well as sharp studies (Egghe & Rao 1992a-b, Egghe & Rousseau 2000), they systematically describe in much detail all that is known about this issue, concluding that "the growth can influence aging but that it does not cause aging".

Well, if we know that growth and obsolescence are closely related, what does the increase in citations to old documents reported in this study means? Does it mean that we are in a period of slow scientific growth? Or what is the same, is science growing exponentially like in previous periods? Or, is today's scientific production of a lower quality, not providing as many new discoveries and techniques? These are all interesting as well as disturbing questions (Bohannon 2014).

In Figure 2 we provide the annual evolution growth of the main three scientific databases today (Web of Science, Scopus, and Google Scholar). The evolution of Microsoft Academic Search is provided as well, though it is hasn't been regularly updated since 2010 (Orduña-Malea, Martín-Martín, Ayllón, Delgado López-Cózar, 2014).

This evolution should be taken with caution since the coverage of each database is different, and the evolution depends on indexing policies. Thus a careful analysis of the exponential growth of scientific literature should be performed.

**Figure 2. Accumulated growth of scientific production (nº of documents) over the years in Google Scholar, Microsoft Academic Search, Web of Science and Scopus.**

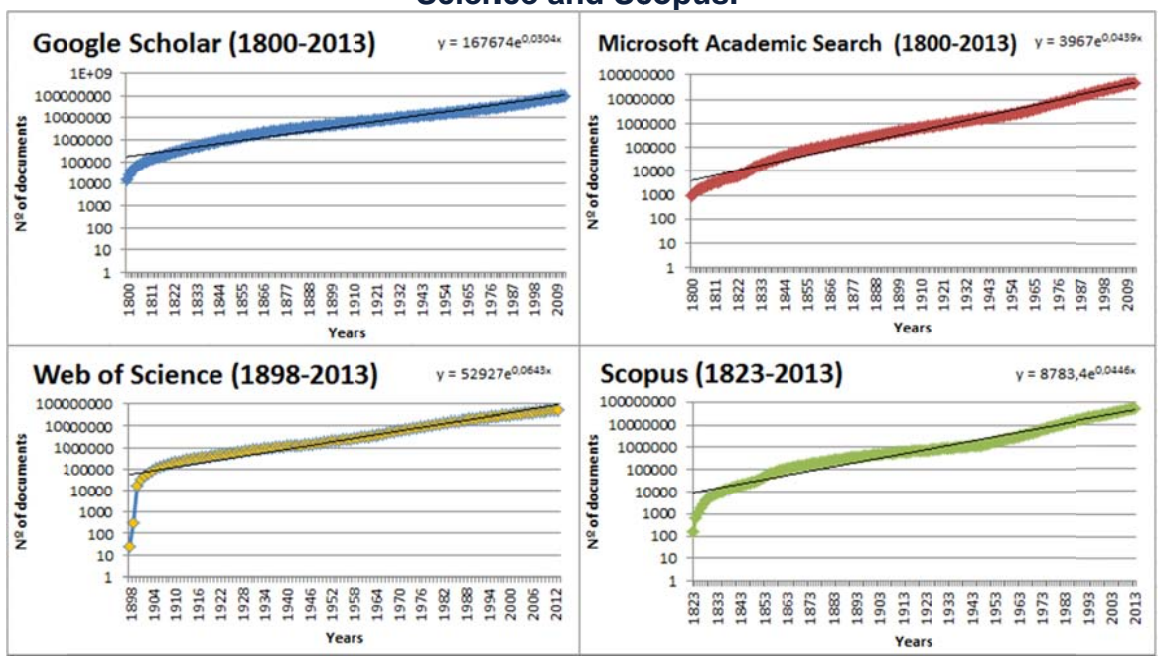

The second factor that may explain the growth in the fraction of citations to old documents, (thus avoiding being quickly forgotten) could be the recent changes in scientific communication brought about by advancements in information and communication technologies (the development of the Internet, the creation and widespread adoption of the Web). The study by Google Scholar's team heads in this direction when they say that widespread citation of older documents is now possible thanks to accessibility improvements to scientific knowledge (digitization of old documents and proliferation of repositories and search engines).

The truth is that these arguments seem quite reasonable, and they are supported by the changes in scientist's reading habits detected by Tenopir & King (2008). For a sample of US science faculties, they find that the advent of digital technologies on searching, storing and publishing has had a dramatic impact on information seeking and reading patterns in science, since scientists have substantially increased their number of readings, read from a much broader range of sources of articles due to access to enlarged library electronic collections, on-line searching capabilities, access to other new sources such as author websites, and, what is more relevant to the issue at hand, "the age of articles read appears to be fairly stable over the years, with a recent increase in reading of older articles. Electronic technologies have enhanced access to older articles, since nearly 80% of articles over ten years old are found by on-line searching or from citation (linkages) and nearly 70% of articles over ten years old are provided by libraries (mostly electronic collections)".

Apart from the effect that improved accessibility (thanks to technological advancement) has on the citations to old documents, the influence of Google Scholar on these changes should also be considered. Since it has already been established that Google Scholar has become the most used source for searching scientific information (Orduña et al. 2014a), that it is the largest public source of scientific information in existence (Orduña et al. 2014b, Ortega 2014), and that it grows at a higher pace than its competitors (Orduña & Delgado López-Cózar 2014), we may conclude that this search engine is contributing in a significant way towards this trend. Although Verstak et al. don't explicitly say it, the use of the product's motto "Stand on the shoulders of giants" in the title of the study might be interpreted as suggesting that the growing trend to cite old documents has been in part caused by Google.

There is truth to this claim, since it is undeniable that Google Scholar has revolutionized the way we search and access scientific information (Van Noorden 2014). A clear manifestation of this is the way results are nowadays displayed in most search engines and databases, a key issue that determines how the document is accessed, read, and potentially cited. The "first results page syndrome", which is causing that users are increasingly getting used to access only those documents that are displayed in the first results pages. In Google Scholar, as opposed to traditional bibliographic databases (Web of Science, Scopus, Proquest) and library catalogues, documents are sorted by relevance and not by their publication date. Relevance, in the eyes of Google Scholar, is strongly influenced by citations (Beel & Gipp 2009, Martín-Martín et al. 2014).

Google Scholar favors the most cited documents (which obviously are also the oldest documents) over more recent documents, which have had less time to accumulate citations. Although it is true that GS offers the possibility of sorting and filtering searches by publication date, this option is not used by default. On the other hand, traditional databases do the exact opposite: trying to prioritize novelty and recentness in their searches (the criterion the have always thought the user will be most interested in) they sort their results by publication date by default, allowing the user to select other criterion if they are so inclined (citation, relevance, name of first author, publication name, etc...).

The question is served. Is Google Scholar contributing to change reading and citation habits because of the way information is searched and accessed through its search engine? If this is true, we could say that the work of the thousands of intellectual laborers that support with their citations the findings of the heroes of science is resting on the shoulders of a GIANT, and that giant has taken the form of a search engine.

**Figure 3. "Standing on the Shoulders of Giants" Wikipedia, 2013. http://en.wikipedia.org/wiki/Standing_on_the_shoulders_of_giants**

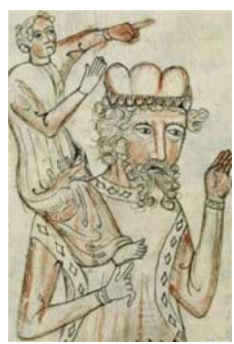